\documentclass[sigconf]{acmart}
\usepackage[utf8]{inputenc}
\usepackage[T1]{fontenc}
\usepackage[normalem]{ulem}
\usepackage{amsmath}
\usepackage{bbm}
\usepackage{xcolor}
\usepackage{natbib}
\usepackage{graphicx}
\usepackage{amsmath}
\usepackage{makecell}
\usepackage{amsfonts}
\usepackage{hyperref}
\usepackage[inline]{enumitem}
\usepackage{booktabs}
\usepackage{multirow}
\usepackage{mathtools}
\usepackage{multicol}
\usepackage{amsmath, bm}
\usepackage{stmaryrd}
\usepackage{longtable}
\usepackage{autobreak}
\usepackage{breqn}
\usepackage{scalerel}
\usepackage{tabu}
\usepackage[ruled,vlined,english,onelanguage]{algorithm2e}
\usepackage{cleveref}
\usepackage[page]{appendix}

\theoremstyle{definition}

\newcommand{\ie}{\emph{i.e.}}
\newcommand{\eg}{\emph{e.g.}}

\newcommand{\vs}{\emph{vs. }}
\newcommand{\reform}{GSQR}

\AtBeginDocument{%
  \providecommand\BibTeX{{%
    \normalfont B\kern-0.5em{\scshape i\kern-0.25em b}\kern-0.8em\TeX}}}

\newif\ifincludeappendix
\includeappendixtrue

\copyrightyear{2023}
\acmYear{2023}
\setcopyright{acmlicensed}\acmConference[SIGIR '23]{Proceedings of the 46th International ACM SIGIR Conference on Research and Development in Information Retrieval}{July 23--27, 2023}{Taipei, Taiwan}
\acmBooktitle{Proceedings of the 46th International ACM SIGIR Conference on Research and Development in Information Retrieval (SIGIR '23), July 23--27, 2023, Taipei, Taiwan}
\acmPrice{15.00}
\acmDOI{10.1145/3539618.3592034}
\acmISBN{978-1-4503-9408-6/23/07}

\begin{document}
\fancyhead{}

\title{Patterns of gender-specializing query reformulation}

\author{Amifa Raj}
\authornote{Work done during internship at Microsoft.}
\affiliation{
  \institution{Boise State University}
   \country{Boise, USA}
}
\email{amifaraj@u.boisestate.edu}
\author{Bhaskar Mitra}
\affiliation{
  \institution{Microsoft Research}
   \country{Montréal, Canada}
}
\email{bmitra@microsoft.com}
\author{Nick Craswell}
\affiliation{
  \institution{Microsoft}
   \country{Bellevue, USA}
}
\email{nickcr@microsoft.com}
\author{Michael D. Ekstrand}
\affiliation{
  \institution{Boise State University}
   \country{Boise, USA}
}
\email{ekstrand@acm.org}

\begin{abstract}

Users of search systems often reformulate their queries by adding query terms to reflect their evolving information need or to more precisely express their information need when the system fails to surface relevant content.
Analyzing these query reformulations can inform us about both system and user behavior.
In this work, we study a special category of query reformulations that involve specifying demographic group attributes, such as gender, as part of the reformulated query (\eg, ``olympic 2021 soccer results'' $\to$ ``olympic 2021 \emph{women's} soccer results'').
There are many ways a query, the search results, and a demographic attribute such as gender may relate, leading us to hypothesize different causes for these reformulation patterns, such as under-representation on the original result page or based on the linguistic theory of markedness.
This paper reports on an observational study of gender-specializing query reformulations---their contexts and effects---as a lens on the relationship between system results and gender, based on large-scale search log data from Bing.
We find that these reformulations sometimes correct for and other times reinforce gender representation on the original result page, but typically yield better access to the ultimately-selected results.
The prevalence of these reformulations---and which gender they skew towards---differ by topical context.
However, we do not find evidence that either group under-representation or markedness alone adequately explains these reformulations.
We hope that future research will use such reformulations as a probe for deeper investigation into gender (and other demographic) representation on the search result page.


\end{abstract}


\maketitle

\section{Introduction}
\label{sec:intro}
Searchers may reformulate their queries to reflect their evolving information needs or in response to less-than-relevant results from the search system to better specify the information they are looking for.
A specific class of query reformulation, often referred to as \emph{specialization}~\citep{jansen2007patterns}, involves a reformulated query on the same topic as the original query but with an expressed intent for more specific information.
Specialization in query reformulation typically involves addition of one or more new terms to the original query~\citep{liu2010analysis, liu2010analysis2}.
In this paper, we are interested in a particular type of query reformulation that involves specialization by adding demographic group attributes, such as gender, to the query.
For example, the query ``NCAA scores'' may be reformulated to ``NCAA \emph{women's}  scores'' to clarify that the searcher is interested in the results for the women's basketball scores at the NCAA; we call these \emph{gender-specializing query reformulations} (\reform{}).
Searchers reformulate their query in response to the results returned by the system which presumably did not fully meet their information need.
Studying such reformulations and the search result pages (SERPs) that led to them may shed light on how that SERP and the results it contains connect to the user's information need, particularly with respect to the specifying term.

In this work, we are interested in situations where users add gender-specifying terms in their query reformulation as a lens on the relationship of system results and gender.
There are many ways a query, the results, and a demographic attribute such as gender may relate.
For example, the linguistic theory of \emph{markedness} \citep{waugh1982marked, bucholtz2001whiteness}---that certain assumptions are assumed in describing a particular class or activity, and markers indicating demographics or other attributes are only used when deviating from the default---leads to an hypothesis that such reformulations may arise when one group dominates the SERP and a different group is desired.

We cannot assume, however, that a \reform{} necessarily means that the specified gender was insufficiently represented in the original SERP, there are many reasons the searcher could reformulate their query, including by clicking on a reformulation the system suggests, or looking to entirely filter out a minority of results that don't meet their needs.
In this initial exploratory investigation of \reform{} behavior, we characterize these reformulations as they appear in a real-world search log from Bing and study their contexts and effects in order to enable future research that can use such reformulations to develop deeper insight into user information need mismatch and search result representations of gender (and other demographics).

To summarize, the key contributions of our current work are:
\begin{enumerate}
    \item We analyze in what context \reform{} occur and factors that may contribute to these reformulation patterns.
    \item We study the impact of said query reformulations on SERPs. 
\end{enumerate}
Finally, we conclude with a discussion of implications of our study on future research and the designe of information access systems.
\section{Related work}
\label{sec:related}




Several research works have analyzed query reformulation to understand various associated aspects including reformulation patterns, applications, and user behavior \cite{hollink2012explaining, mitraexploring, jansen2007patterns}. Previous research identified that users often add terms (\textit{specialization}) or remove terms (\textit{generalization}) to modify their queries \citep{bruza1997query, boldi2011query, jansen2009patterns}. This behavior of query modification is found to be effective in retrieving more relevant information for users \cite{gauch1993expert, rieh2006analysis}. Several studies worked towards identifying patterns of user query reformulation behavior, and these patterns were categorized based on search task, sequences, user intent, or semantic analysis \cite{huang2009analyzing, rieh2006analysis, liu2010analysis2}. \citet{huang2009analyzing} provided a taxonomy of users' query reformulation strategies by focusing on how users write reformulations. For example, users may add, remove, substitute, or reorder to reformulate their queries. \citet{liu2010analysis} analyzed how users reformulated queries for different search tasks and found that the type of search task can effect their reformulation behavior. \citet{liu2010analysis2} tried to understand the connection between task types, SERP of previous search and users' query reformulation behavior. Moreover they categorized query reformulation type into five groups: generalization, specialization, term substitution, repeat, and new, and further observed the effectiveness of their reformulation regarding different search tasks. \citet{rha2017exploration} and \citet{chen2021towards} analyzed user behavior regarding different query reformulation techniques to identify user intention behind their reformulation. These studies observed that users may reformulate queries to find specific results, learn more about a topic, or satisfy their particular needs. This knowledge of query reformulation patterns and user intention associated with the reformulation further facilitate research on designing search engines that can better support user information need \cite{boldi2011query, chen2019investigating}.
In our work, we are specifically examining specialization reformulations that specify a demographic group (gender) as a lens to understand users' intent and system responses to both initial and refined queries in such settings.

\section{Data and methods}
\label{sec:method}

\paragraph{Identifying \reform{} in search logs.}
Our study focuses on query reformulation patterns that specializes the information need to specific gender groups.
We adopt a narrower definition of specializing query reformulations compared to Liu et al.~\citep{liu2010analysis, liu2010analysis2}.
We define a query reformulation to be specializing if the reformulated query contains all the terms in-order as in the original query and includes an additional set of contiguous terms added anywhere to the original query.
We enforce a constraint that the additional terms must include one term from a list of known gender terms---\ie, ``woman'', ``man'', ``women'', ``men'', ``woman's'', ``man's'', ``women's'', ``men's'', ``womans'', ``mans'', ``womens'', ``mens'', ``female'', ``male'', ``male's'', ``female's'', ``males'', and ``females''---and optionally additional terms from a known list of prepositions---\ie, ``about'', ``against'', ``according to'', ``among'', ``at'', ``by'', ``except'', ``for'', ``from'', ``in'', ``like'', ``of'', ``on'', ``to'', ``with'', and ``without''.
In line with this definition, we consider reformulations like ``leadership quotes'' $\to$ ``leadership quotes by women'' and ``bmi calculator'' $\to$ ``bmi calculator \emph{for men}'' as \reform{}.
Given large-scale search logs we can automatically identify instances of query reformulations matching these specified patterns. 
We understand the risk of considering gender as binary attribute \cite{pinney2023much}; our analyses can be further extended to non-binary gender and other demographic attributes.
\ifincludeappendix
To consider other demographic groups beyond gender, it would be convenient if we can (semi-)automatically detect other relevant group terms. Please see Appendix~\ref{sec:appendix-extend} for more details.
\fi
\paragraph{Data.}
To understand user intentions behind \reform{}, we analyze instances of similar reformulations in large scale search logs over the period of one year (January 1 -- December 31, 2021) from Bing.
We note that although specialization is a common form of reformulation~\citep{jansen2007patterns, jansen2009patterns, liu2010analysis, liu2010analysis2}, people enter a vast variety of specializing reformulation terms, and our gender terms are only a small fraction.
From the search logs, we extract a sample of approximately $4.7$ million pairs of consecutive queries from the same search session where the second query is a \reform{} of the former. This was 3.9\% of the specializing queries we considered.
For both original and the reformulated queries, we extract metadata such as timestamps, entry point from which the query was submitted, web results that were displayed to the user, and a record of user clicks (if any) on those results.

\paragraph{Analysis methods.}
The focus of our study is to understand user intent behind \reform{} and with that goal we did following analyses:
\begin{itemize}
    \item To ensure that this is a recurring pattern, we consider the frequency of such reformulations in our search log data.
    \item We analyze how the pattern differs across query topics by using an automated text-based classifier on the original queries.
    \item We analyze time differences between original and reformulated queries and potential elements on the original SERP---\eg, related query recommendations for intent disambiguation---that may influence the user to reformulate their queries.
    \item We analyze which groups are specified more often in aggregate in these reformulations by using average GloVe embeddings~\citep{pennington2014glove} of terms as query representation and based on the approach proposed by \citet{bolukbasi2016man}, we compute a genderedness measure for queries.
    \item We study the impact of these reformulations on the SERP.
\end{itemize}

\section{Results}
\label{sec:analysis}

\begin{table}[]
    \caption{Specialization patterns by topic. \reform{} rate is compared to all-topic average. \% rec. and \% woman are percent of \reform{} that were clicked recommendations and added woman gender terms, respectively.}
\footnotesize
\begin{tabular}{lrrr}
\toprule
Original query topic &   \reform{} rate & \% rec. & \% woman \\
\midrule
shopping and fashion &  15.7 &    13\% &     50\% \\
health               &   2.0 &    23\% &     52\% \\
sports and outdoors  &   2.0 &     7\% &     62\% \\
parenting            &   1.7 &    17\% &     41\% \\
animals              &   1.2 &    19\% &     56\% \\
psychology           &   1.1 &    26\% &     54\% \\
religion             &   0.9 &    17\% &     66\% \\
art                  &   0.8 &    11\% &     56\% \\
literature           &   0.7 &    18\% &     70\% \\
philosophy           &   0.7 &    16\% &     52\% \\
history              &   0.6 &     4\% &     82\% \\
photography          &   0.5 &    12\% &     60\% \\
entertainment        &   0.4 &    11\% &     55\% \\
other                &   0.4 &     9\% &     51\% \\
science              &   0.3 &     8\% &     49\% \\
cooking and food     &   0.3 &    11\% &     46\% \\
politics             &   0.3 &     5\% &     70\% \\
travel               &   0.3 &     9\% &     67\% \\
education            &   0.3 &     7\% &     67\% \\
vehicle              &   0.2 &     4\% &     67\% \\
home and garden      &   0.2 &     8\% &     50\% \\
technology           &   0.2 &    15\% &     52\% \\
finance              &   0.2 &     7\% &     57\% \\
\midrule
all topics           &   1.0 &    13\% &      54\% \\
\bottomrule
\end{tabular}    \label{tab:topic_statistics}
\end{table}

\subsection{\reform{} patterns overall and by topic}

\paragraph{Overall Pattern.} As described in Section~\ref{sec:method}, each of our 4.7 million \reform{} cases adds at least one term from our list of gender terms. Of those, 54\% add women-related and 46\% add men-related terms. The median time between the first and second query is 19 seconds. Reformulations that add men-related terms tend to happen slightly more quickly, with a median of 17 seconds, compared to 19 seconds for those adding women-related terms.

The most common methods for entering the second query are by editing at the top of the SERP (64\% of our \reform{} cases with median time 15 seconds) or clicking a recommended query at the top of the SERP, as in \citep{zamani2020analyzing} (13\% of our \reform{} cases with median time 13 seconds).
Clicking a recommended query at the top of the SERP contributes slightly higher for query reformulations specializing \emph{for women} ($14.3\%$) compared to \emph{for men} ($11.5\%$). Besides editing and clicking at the top of the SERP, other methods of \reform{} may involve clicking on other query suggestions on the page or re-entering the query via other entry points (median time 60+ seconds).

So far there are some small differences between men-related and women-related \reform{}, with men-related reformulation added more quickly, and women-related reformulation happening more often and having slightly higher association with recommended queries. 
\paragraph{Query Topic Analysis.} Table~\ref{tab:topic_statistics} breaks down the analysis by topical category of the original query using a topical classifier. The rate column reveals how much more prevalent \reform{} is for that topic than expected, given how often we see that topic in Bing logs.
The rate is particularly high in the \textit{shopping and fashion} category and lower in categories such as \textit{home and garden}, \textit{technology}, and \textit{finance}.

The next column is the percent of \reform{} that came from the query recommendation feature. The rate and recommender columns are correlated (Spearman $\rho$ 0.586, $\mathrm{p-val=0.003}$), suggesting the possibility that reformulation patterns and recommenders drive each other to some extent. It is clear that the rate is not entirely driven by the recommender, because then the \% recommended column would need to explain the ratio of 15.7 for \textit{shopping and fashion}, whereas the actual percentage is 13\%, close to the overall average.
The largest usage of the recommender is in the \textit{psychology} and \textit{health} categories.

The last column shows what percentage of \reform{} are women-related, by topic. For the \textit{shopping and fashion} topic we see a 50\% split, indicating that both men and women need to specify gender in such scenarios. Women-related reformulation is lower for the parenting topic, with 41\%, perhaps suggesting that finding men-related information is taking a bit more effort in this area.
For the history topic, the average fraction of women-related reformulations is 82\%, for example there were two instances of ``on this day in history'' followed by ``on this day in history women''.

\begin{figure}
    \centering
    \includegraphics[width=0.65\linewidth]{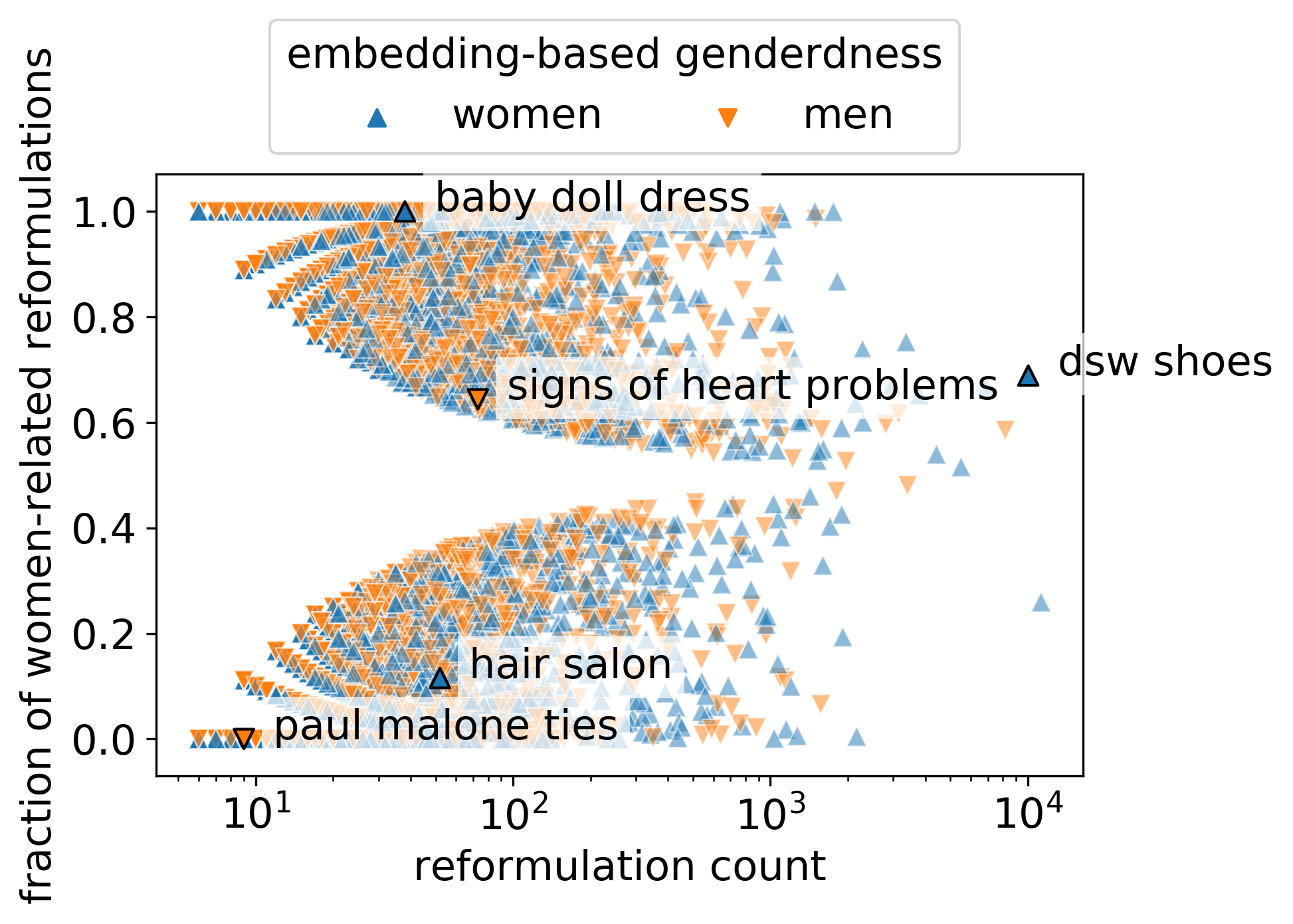}
    \caption{Scatter plot of queries that exhibit genderedness both in reformulation (y-axis) and GloVe embedding (marker). We see both types of embedding-based genderedness receiving all types of reformulation. Four example queries are highlighted.}
    \label{fig:scatter_binomial}
\end{figure}

\begin{figure}[tbp]
    \centering
    \includegraphics[width=0.85\linewidth]{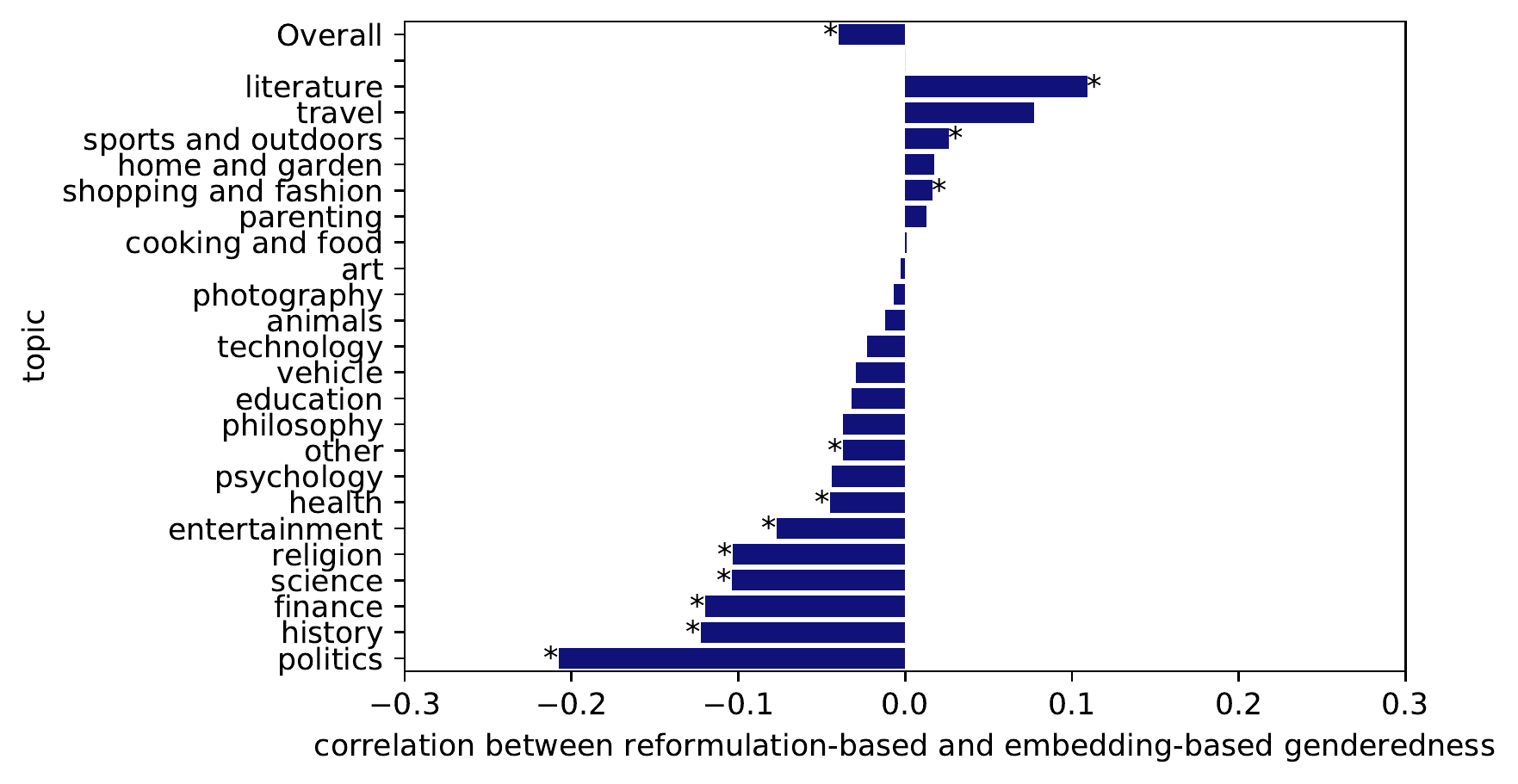}
    \caption{Spearman correlation between reformulation-based and embedding-based genderedness, with statistically significant correlation indicated by * (p-val < 0.05).}
    \label{fig:correlation_by_topic_barplot}
\end{figure}

\subsection{Embedding-based genderedness}

We now consider embedding-based genderedness~\cite{bolukbasi2016man} of the original query, using GloVe. This allows us to study the relationship between the \% women reformulation pattern and the embedding-based genderedness. For example, corresponding to the theory of markedness mentioned, we may see more men-related reformulations if the original query is associated with women (\eg, ``nurse'' $\to$ ``male nurse'').

In Figure~\ref{fig:scatter_binomial}, each point indicates a query. Queries where the reformulation patterns were balanced between men and women were removed, according to a two-tailed binomial test ($\mathrm{p-val < 0.05}$). This gives the plot its funnel shape, since a reformulation with fewer observations (x-axis) has to be extreme on the y-axis for the binomial test to keep it. Also, queries where the embedding-based genderedness was close to 0 (within $\pm 0.05$) were removed, removing about half the queries, and keeping the most women-related queries as blue triangles and the most men-related queries as orange triangles.

The figure shows that there are many queries where the embedding-based genderedness is being corrected by reformulation, but also many where the genderedness is reinforced by the reformulation. Five queries are highlighted as examples. For ``baby doll dress'' and ``paul malone tie'', the genderedness is reinforced by the reformulation pattern, whereas for ``hair salon'' and ``signs of heart problems'', the genderedness of the query is corrected by the query pattern. We also include one query with high reformulation count ``dsw shoes''.


Seeing a great variety of reformulation patterns, which can contradict or reinforce original query's genderedness, raises the question of whether query genderedness and reformulation patterns are correlated overall. Figure~\ref{fig:correlation_by_topic_barplot} shows the overall correlation and the per-topic correlation where all the correlations are low, indicating that we do see a mix of user behaviors, as depicted in Figure~\ref{fig:scatter_binomial}. Over all topics (``Overall'') the correlation is mildly negative, and statistically significant, meaning that an original query having a higher value on fraction of woman-related reformulations is associated with having a lower value on woman-related genderedness. The correlation is weak, but happens more often than would happen by chance. When the correlation is statistically significant ($\mathrm{p-val < 0.05}$) the figure shows it with a *. This means that the reformulations in \textit{literature}, \textit{sports} and \textit{shopping} topics are more likely to reinforce the genderedness of the original query and this pattern is present Figure~\ref{fig:scatter_binomial}, in the dress and ties example queries. For several topics the correlation is the other way, with reformulations tending to correct the genderedness of the original queries. Although we see several statistically significant cases of reinforcing or correcting genderedness, in all cases the associations are mild, with correlation in the range -0.2 to 0.2, so overall the user behavior is mixed, for all topics and overall.





\subsection{Impact of reformulations}
In our query reformulation data, we find several examples via manual inspection where the addition of gender terms surfaces more gender-specific results (\eg, ``ADHD symptoms'' $\to$ ``ADHD symptoms for women'') or effectively corrects for under-representation in the original SERP (\eg, ``US open golf 2022'' $\to$ ``US open golf 2022 women''). If these reformulations are effective then we would expect behavioral search satisfaction metrics, like clickthrough rate (CTR), to improve for the reformulated query SERP in aggregate.
While we cannot disclose exact CTR due to Bing disclosure limitations, we note that reformulated query SERP CTR is $2.6$ times higher than original query SERP CTR.
This ratio is approximately the same for both men and women as the specializing term.
If we consider only queries where the reformulation was through a clarifying query recommendation, the CTR boost increases to $3.6$.

\begin{figure}
\includegraphics[width=0.6\columnwidth]{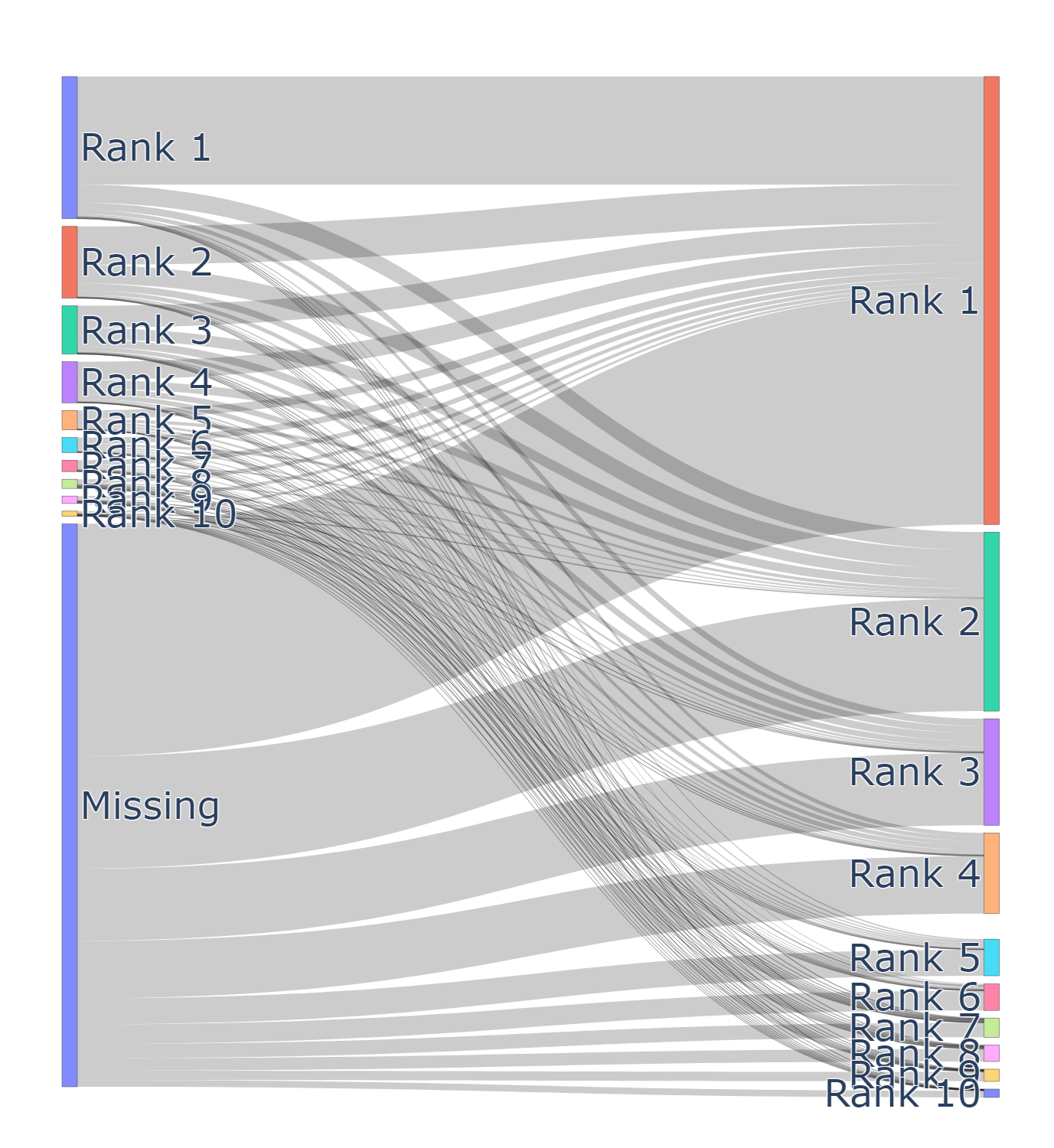}
\caption{Visualization of the change in the rank of a result (clicked on the latter SERP) from original query SERP (left) to reformulated query SERP (right) in cases.
The clicked document is often missing from the original SERP, but in several cases it was also on the original SERP, possibly near the top.}
\label{fig:sankey}
\end{figure}

CTR does not tell the whole story because uses may click on a link in the second SERP because they reformulated quickly and did not carefully examine the first SERP.
We therefore examine the difference the query reformulation makes in the rank of the selected item.
Figure~\ref{fig:sankey} visualizes this analysis; each data point is \reform{} event where the user clicks on a document on the reformulated query SERP. 
In $62\%$ of cases, the clicked result did not appear on the original query SERP.
In another $18\%$ of cases, the clicked document appeared on the original SERP but at a lower rank, and $14\%$ of the time they appear at the same position on both SERPs.
This shows that the reformulated queries are yielding better access to the ultimately-selected results.

\reform{} may also influence the relative visibility of different content sources, especially if publishers have optimized their content for particular queries.
To analyze this potential phenomenon empirically, we compute the ratio of the probability of exposure for individual websites on the reformulated query SERP vs. the original SERP, which we refer hitherto as \emph{exp-ratio}.
We estimate these probabilities with the Expected Exposure technique of \citet{diaz2020evaluating} using the NDCG user behavior model.
As we may expect, websites that contain group-specific content---\eg, \textsf{menshairstylestoday.com} (exp-ratio=$3.7$) and \textsf{menshealth.com} (exp-ratio=$2.3$) for men and \textsf{womenshealthmag.com} (exp-ratio=$2.8$) for women---are exposed significantly more on the reformulated query SERP.
Websites may also gain more exposure on the reformulated query SERP if they specifically create pages for different groups either as part of better content organization or as a search engine optimization technique.
This may sometimes lead to undesirable outcomes such as underexposure of authoritative websites like \textsf{mayoclinic.org} (exp-ratio=$0.8$) and \textsf{webmd.com} (exp-ratio=$0.7$) compared to sites that are better-optimized specifically for such query patterns.


\section{Discussion and conclusion}
\label{sec:conclusion}
In this study we find that users of search systems may reformulate their queries to look for gender-specific results, often to correct for mismatch of the their information need. Our analysis here is a first-pass approach to the problem. There can be multiple factors that can lead users to do \reform{} and by looking at the reformulated queries and how genders are represented on overall SERP---which may include images, videos, and news results---we can develop a deeper understanding of the circumstances that led to these reformulations. 

For example, we identified reformulations that seek \textit{gender-specific information} (e.g., ``ADHD symptoms'' $\to$ ``ADHD symptoms for women'') or reformulations that \textit{intensify group representation} (e.g., ``hispanic names'' $\to$ ``hispanic names for women'') or \textit{reinforce over-representation} (e.g.,  ``NCAA basketball score'' $\to$ ``NCAA men's basketball score'') where users may want to filter-out results related to another gender. We found reformulations that \textit{correct for group under-representation} where users find fewer results about the particular gender relevant to their information need and so may explicitly reformulate the query mentioning that gender to find more results (e.g., ``NCAA basketball score'' $\to$ ``NCAA women's basketball score''). There can be reformulation that are \textit{influenced by other SERP elements} such as, images or videos (e.g., ``hiking boots'' $\to$``hiking boots for women'' because images in original query SERP are skewed towards one ). And lastly, there can be \textit{harmful reformulations} where the reformulated query contains names of people and gender identities.
Many of these queries are harmful speculations about the subject's gender and often misgender the subject.
We intentionally do not include any examples of such queries to avoid perpetuating further harm.
\ifincludeappendix
Please see Appendix~\ref{sec:appendix-manual} for additional manual analysis of GSQRs.
\fi

Through our study, we show the importance of analyzing \reform{} to get deeper insight about user behavior and systems. Trace ethnography~\citep{geiger2011trace} combines analysis of data traces from large-scale online systems, 
with ethnographic techniques to deeply understand the journeys users take in their use of the system. Query reformulations seem likely to be a useful lens to focus on such studies---applying trace ethnography to sessions that contain demographic reformulations, for instance, may produce a richer understanding of users' search goals and behavior and the system's response.
As biases in group representations in search results may contribute to social harms and unfairness, more of such studies should be conducted that may shine a light on disparities in group representation and also provide insights that may be instrumental in developing measures of representational bias and representational
\ifincludeappendix
harms~\citep{crawford2017trouble}, although there are additional challenges in measuring representational bias and fairness as we discuss in Appendix~\ref{sec:appendix-fairness}.
\else
harms~\citep{crawford2017trouble}.
\fi
Such studies may be complemented by other forms of inquiry, such as lab-based user studies and online surveys, to further elicit situations where group representations are important to consider in the context of online information access.
Understanding implications of how groups are represented in retrieved information may have important implications for designs of future information access systems.
We hope that future research continues to engage with these questions of moral import.
~\nocite{mitraexploring, agrawal2002indigenous, crawford2021atlas, chasalow2021representativeness, miceli2022studying, sen2008idea, sambasivan2021re, wong2012dao, sweeney2013discrimination, jacobs2021measurement}

\small{
\bigskip\noindent
\textbf{Acknowledgments.}
This work was partially supported by the National Science Foundation under grant IIS 17-51278.
}
\balance
\bibliographystyle{ACM-Reference-Format}
\bibliography{references}

\ifincludeappendix
\begin{appendices}
\section{Extending to other demographics}
\label{sec:appendix-extend}
In our work, we focus primarily on analyzing query reformulations corresponding to gender-based specialization.
However, similar analysis would also be meaningful considering other demographic attributes, say race and age, and intersectional identities---\eg, ``black women'' and ``elderly man''.
To extend our analysis beyond gender, it would be convenient if we can (semi-)automatically detect other relevant group terms.
Towards that goal, we represent every specializing query reformulation in terms of a template and a keyphrase.
For example, for the query reformulation ``hairstyles'' $\to$ ``hairstyles for women over 50'', we identify the template to be ``hairstyles for [KEYPHRASE]'' and the corresponding keyphrase as ``women over 50''.
Obviously, this applies even to specializing query reformulations that do not strictly correspond to demographic groups---\eg, for the query reformulation ``durable shoes'' $\to$ ``durable shoes for hiking'' we identify ``durable shoes for [KEYPHRASE]'' as the template and ``hiking'' as the keyphrase.
Let $T$ be the set of templates and $K$ the set of keyphrases extracted from a search log.
Furthermore, let $K_{\text{seed}} \subset K$ be a set of seed keyphrases that corresponds to demographic group attributes---\eg, based on gender.
Then, to identify other keyphrases that correspond to demographic group attributes in this context, we score each keyphrase $k \in K$ as follows:

\begin{align}
    S_k &= \sum_{t \in T}{p(k|t) \cdot p(t|K_{\text{seed}})}
\end{align}

We rank the candidate keyphrases in descending order based on this score and manually inspect the top candidates to identify the keyphrases that correspond to demographic group attributes.
Using this method on a single day of search logs from a commercial web search engine and a seed set of gender-related group terms, we are able to identify the following additional group terms in the top dozen keyphrases: ``kids'', ``girls'', ``boys'', ``teens'', ``seniors'', and ``women over 50''.
In addition to inspecting the keyphrases listed lower down in this ranklist, more group terms may be discoverable by employing this approach iteratively to grow the seed set of group terms and by running the analysis over a search log corresponding to a larger time period.
In future work, it would be interesting to also explore other approaches, such as considering latent representations of query reformulations~\citep{mitraexploring}, for identifying new group terms and corresponding group-based specializing query reformulation patterns.
\section{Manual analysis}
\label{sec:appendix-manual}


Through our manual analysis, we want to get insight on how gender is represented on SERPs and user intent behind their group-based query reformulation.
By manually verifying how genders are represented on overall SERP---which may include images, videos, and news results---we aim to develop a deeper understanding of the circumstances that led to these reformulations.
We manually analyze a sample of reformulations from the search log data which leads us to the following additional insights about user query reformulation behaviors in this context:

\paragraph{Reformulations that seek gender-specific information.} 
We find examples where the original query SERPs do not necessarily show any gender indicators through images, videos, or snippets of web results and users are specifically looking for gender specific information.
For example, when users search for ``ADHD'' symptoms, the original query SERP does not show any gender specific results; there is no mention of gender specific terms in the snippets of retrieved web results. Users reformulate their query for particular gender to get gender-specific ADHD symptoms. Even though the original query SERP looks gender neutral and relevant to both genders, users may want to learn more about that topic for a specific gender. In the reformulated query SERP for ``ADHD symptoms for women'', we notice more gender-specific results. The first result has the term ``women'' in it and it is the women specific section from add.org which is the official website of Attention Deficit Disorder Association. Moreover, almost all of the web results contain the term ``women'' and all image and video results also contain content relevant specifically to women.

\paragraph{Reformulations that intensify group representation or reinforce over-repr\-esentation.}
In this scenario, users may want to filter-out results related to another gender and only want results regarding the specified gender in their reformulated query.
Looking at examples, we found SERPs where results are targeted for both genders. For example, where users search for ``hispanic names'', original query SERP shows results that are relevant for both men and women but users want gender specific hispanic name like ``hispanic names for women''. In reformulated query SERP, we notice the images and videos become more gender specific.
We also notice that the webpage ``100 Gorgeous Hispanic Girl Names'' appears in both SERPs but at 3rd position in original query SERP and is located in the 1st position in reformulated query SERP. 

We found queries where the original query SERPs are slightly skewed towards one gender through it's content; particular gender is slightly more dominant in image and video representations and texts in retrieved web pages. 
For example, the original SERP of the query ``cordless razor'' has more mention of men in retrieved web pages and the images are mostly targeted to men. However, the next query is ``cordless razor for men'' which emphasizes on men-specific results and exclude gender-neutral results. In this scenario, reformulated query SERP shows more gender-specific results; there are more web pages on reviews of cordless razor for men. 
In both SERPs, the first results are from amazon, however in the reformulated query SERP, the link is directing specifically to cordless razors for men.

There are queries where original SERPs show notable differences in gender representation through their content including images, videos, news, and texts in retrieved web pages. Such unequal representation of gender in SERP can happen due to reflecting societal stereotypes or bias. For example, the SERP of the original query ``NCAA basketball score'' shows results mostly about men's team, games, and game scores; images, news, and videos also heavily represent men. When user reformulate their queries to ``NCAA men's basketball score'', they may want more results related to men and exclude any gender-neutral results. In this scenario, the reformulated query SERP does not change a lot from the original query SERP; this shows more results that are relevant to men.

\paragraph{Reformulations that correct for group under-representation.}
In this scenario, users find fewer results about the particular gender relevant to their information need and so may explicitly reformulate the query mentioning that gender to find more results.
Original SERP slightly over-represent one gender through images and texts. For example, the original SERP for the query ``dillards shoes'' shows products related to women through images and ads and thus the results seem more relevant and targeted to women. The next query is ``dillards shoes for men'' which brings results suitable and relevant for men. In this scenario, reformulated SERP shows differences in image and videos than the original SERP. The top images of the original SERP has more women's shoes but in reformulated query SERP all of the images were for men. The first result of the reformulated query SERP in the men's section from the official site of Dillard's which did not appear in original query SERP.

Original SERPs show notable differences in gender representation in retrieved images, videos, news, and webpages. For example, the original SERP of the query ``NCAA basketball score'' shows results mostly about men's teams, games, and game scores; images, news, and videos also heavily represent men. Only one retrieved web page is about women in original query SERP. Hence, the next query is often ``NCAA women's basketball score'' where users can find results relevant and more specific to women. In this scenario, reformulated SERP shows visible differences than original SERP including images, videos, and news.

\paragraph{Reformulation that are influenced by other SERP elements.}
In some cases, the web results and other results on SERP may represent groups differently.
For example, the original query SERP of ``hiking boots'' shows web pages that are relevant and suitable for both men and women. However, the images are mostly targeted for men. Even though the web pages may show gender neutral results, images are skewed towards one gender. When users reformulate their query to ``hiking boots for women'', they may want to see SERP which looks visually more gender-specific and relevant to their gender intent.

\paragraph{Harmful reformulations.}
During our manual analysis, we also come across few cases where the reformulated query contains names of people and gender identities.
Many of these queries misgender the subject and in fact these queries are likely outcome of harmful speculations about the subject's gender.
We intentionally do not include any examples of such queries to avoid perpetuating further harm.
\section{On the challenges of measuring representational bias and fairness}
\label{sec:appendix-fairness}
In this study, we analyze user behavior of searchers to gain more insights about how different demographic groups are represented in search results.
Our findings indicate complex interactions between group representations on SERPs and other social factors contribute to the observed query reformulation patterns.
We hope that this current work serves as a step towards understanding group representations in search results, and subsequently towards developing appropriate formalization for representational bias and fairness.
However, there are additional challenges in making progress towards formalizing representational bias and fairness that future work should be wary of that we discuss in this section.

\paragraph{Politics of classification.}
A particularly difficult challenge for bias and fairness analysis is associating demographic group labels with the results returned by the search system.
This goes far beyond issues of developing and operationalizing appropriate taxonomies for annotation, which in itself raises challenging questions, such as how to annotate artefacts that may not include any demographic markers or includes content corresponding to multiple groups, and whether the semantics of the group labels should reflect who the content is for, about, or by.
But a more fundamental issue is that any classification, and corresponding development of taxonomies, are inherently political actions~\citep{agrawal2002indigenous, crawford2021atlas}.
Even our own analysis in this study suffers from many of these issues---for example, focusing on two genders perpetuates the false and problematic framing of gender as a binary construct and even with acknowledgment of that fact continuing to conduct analysis and present results within that binary frame contributes to erasure of other gender identities.
Attempts to find proxy measures that can be used to automatically assign group labels (such as gender) to results takes away the subjects' rights to self-identify and may result in their misgendering.
Finally, any automatic scheme for annotating results with demographic attributes may also find harmful applications in the hands of nefarious actors and requires that as researchers we handle these questions with utmost caution and necessary sensitivity and thoughtfulness.

\paragraph{What is fair?}
Adopting a normative view of fairness in the context of information access systems is at best operationally challenging and more likely in itself problematic.
For example, let us consider the normative position that a search system is responsible for appropriate representation of all groups on the search result page.
Leaving aside the challenges of defining what constitutes representativeness~\citep{chasalow2021representativeness} in this context, the crude framing in which we associate group labels with individual results and attempt to achieve some ``fair'' target distribution on the result page leads us to a fundamental question: How do we choose a fair target distribution?
Let us, for example, consider the question of representation of content corresponding to men's and women's sporting events on the search result page.
Is the correct target distribution that both types of content should be exposed equally? Or, should it reflect the proportion in which these sporting events occur in the real world (which due to historical gender discrimination may skew it towards a specific gender)? Or, should it reflect the proportion of online content corresponding to these events (\ie, supply-side distributions which may further skew it towards a specific gender)? Or, should it model the proportion of search queries corresponding to these events (\ie, demand-side distribution which may again be disproportionately skewed towards a specific gender)? In reality, the target distribution may also be influenced directly or indirectly by the commercial interests of the owner of the search system (\eg, based on monetizability of content) which is even more problematic.
These choices require thoughtful consideration and as system designers it is less than obvious what ``fair'' distribution the systems should aim for.
In fact, whether fairness is the correct framing for these problems should itself be questioned~\citep{miceli2022studying}, and we should probably also look past the cultural hegemony of our own research communities to look for alternative framings for harm reduction~\citep{sen2008idea, sambasivan2021re, wong2012dao}.

\paragraph{Harmful exposure.}
Not all exposure is desirable.
In the context of representational harms, it is crucial to consider the context and the way a subject may be exposed.
Examples of undesirable exposure include when searches for black identifying first names in online ad delivery systems disproportionately suggest arrest record searches~\citep{sweeney2013discrimination}.
Representational harms also occur when an otherwise regular search query in the context of specific demographic groups return harmful stereoptypical or sexualized caricatures---\eg, the queries ``asian men'' \vs ``asian women''.
Any measure of representational harm must therefore not only consider the amount of exposure a group receives but in what context and form.

\paragraph{Measurement validity}
Finally, any measure of bias, unfairness, or harm must be rigorously tested for construct validity, especially for consequential validity~\citep{jacobs2021measurement}.
What are the short-term and long-term consequences of optimizing search systems towards said measures?
Are they likely to reduce harm or compound them, including in ways that may not be immediately obvious or easy to measure?

\end{appendices}
\fi

\end{document}